\documentclass[lettersize,journal]{IEEEtran}
\usepackage{amsmath,amsfonts}
\usepackage{algorithmic}
\usepackage{algorithm}
\usepackage{array}
\usepackage[caption=false,font=normalsize,labelfont=sf,textfont=sf]{subfig}
\usepackage{textcomp}
\usepackage{stfloats}
\usepackage{url}
\usepackage{verbatim}
\usepackage{graphicx}
\usepackage{cite}
\usepackage{xcolor}
\usepackage[utf8]{inputenc}
\usepackage[T1]{fontenc}
\usepackage{multirow}      
\usepackage{booktabs}      
\usepackage{threeparttable} 
\usepackage{graphicx}
\usepackage{array}
\begin{document}

\title{NLS: Natural-Level Synthesis for Hardware\\ Implementation Through GenAI}

\author{Kaiyuan Yang, Huang Ouyang, Xinyi Wang, Bingjie Lu, Yanbo Wang, Charith Abhayaratne,~\IEEEmembership{Member,~IEEE,}\\ Sizhao Li,~\IEEEmembership{Member,~IEEE,} Long Jin,~\IEEEmembership{Senior Member,~IEEE,} Tiantai Deng
\thanks{This work was supported by the AI for Productive Research \& Innovation in eLectronics (APRIL) Hub under Grant XX/XXXXXXX/X.}
\thanks{Manuscript received February XX, 2025; revised August 16, 2021.}}

\markboth{IEEE Transactions on Computer-Aided Design of Integrated Circuits \& Systems,~Vol.~XX, No.~X, February~2025}%
{Shell \MakeLowercase{\textit{et al.}}: NLS: Natural-Level Synthesis for Hardware Implementation Through GenAI}

\IEEEpubid{
\makebox[\columnwidth]{0000--0000/00\$00.00~\copyright~
2025 IEEE \hfill}
\makebox[\columnwidth]{}
}

\maketitle
\IEEEpubidadjcol

\begin{abstract}
This paper introduces Natural-Level Synthesis (NLS), an innovative approach for generating hardware using generative artificial intelligence (Gen-AI) on both the system level and component-level. NLS bridges a gap in current hardware development processes, where algorithm and application engineers’ involvement typically ends at the requirements stage. With NLS, engineers can participate more deeply in the development, synthesis, and test stages by using Gen-AI models to convert natural language descriptions directly into Hardware Description Language (HDL) code. This approach not only streamlines hardware development but also improves accessibility, fostering a collaborative workflow between hardware and algorithm engineers. We developed the NLS tool to facilitate natural language-driven HDL synthesis, enabling rapid generation of system-level HDL designs while significantly reducing development complexity. Evaluated through case studies and benchmarks using Performance, Power, and Area (PPA) metrics, NLS shows its potential to enhance resource efficiency in hardware development. This work provides a extensible, efficient solution for hardware synthesis and establishes a Visual Studio Code (VS Code) Extension to assess Gen-AI-driven HDL generation and system integration, laying a foundation for future AI-enhanced and AI-in-the-loop Electronic Design Automation (EDA) tools. 
\end{abstract}


\section{Introduction}
\IEEEPARstart{H}{ardware} is the key to support computing-intensive applications like Artificial Intelligence (AI), Digital Signal Processing (DSP), and image processing \cite{ref1,ref2,ref3}. New hardware architectures from industry and academia, such as the Google Tensor Processing Unit (TPU), Nvidia A100/H200 Graphics Processing Unit (GPU) and Field Programmable Gate Array (FPGA)-based accelerators, are keeping pace with the rapidly growing computational demands of algorithms \cite{ref4,ref5,ref6}. From humble silicon wafers to powerful supercomputers, hardware forms the cornerstone of advanced technology. As the demand for more complex and efficient hardware grows, making development processes quicker and more efficient becomes increasingly important, especially to shorten the time to bring the product to the market or publications in academia. 

From a designer's perspective, hardware development is becoming more complicated, while development tools are evolving to enforce greater logical precision and efficiency. Hardware development methodologies have progressed from detailed gate-level design to more abstract Register Transfer Level (RTL) and behavioural levels, as previous methods became inadequate for modern hardware complexity \cite{ref7}. High-level language tools have emerged to simplify the development process significantly, addressing the growing complexity and enabling more efficient design workflows \cite{ref8}. These ongoing improvements enhance efficiency and expand access to hardware development, allowing professionals such as AI algorithm and application engineers to engage in hardware design previously beyond their expertise.

High-Level Synthesis (HLS) tools simplify hardware development by converting High-Level Languages (HLL) into Hardware Description Languages (HDL). Examples include Vitis HLS for C/C++, MyHDL for Python, LabVIEW FPGA module for LabVIEW, and HDL Coder for MATLAB \cite{ref9,ref10,ref11,ref12}. These tools simplify design by hiding hardware complexities, allowing engineers to use familiar HLL syntax at a higher abstraction level than traditional RTL coding. These advancements have made hardware development more accessible and efficient.

The hardware development process is crucial for ensuring the efficiency and reliability of computing systems and devices \cite{ref13}. As shown in Fig. 1, this process typically begins with the Requirement and Specification phase, where algorithm and application engineers define specific needs and expectations for the hardware.

\begin{figure}[!t]
\centering
\includegraphics[width=3.5in]{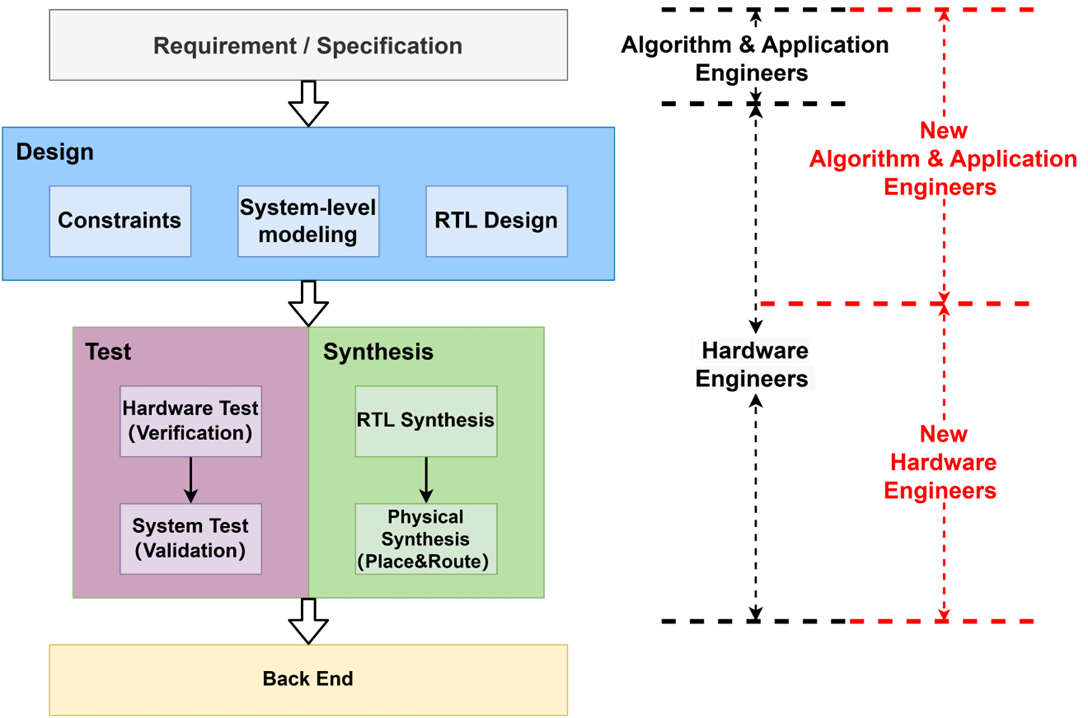}
\caption{Hardware development process.}
\label{fig_1}
\end{figure}

Next is the Design phase, primarily handled by hardware engineers. They begin by setting constraints and then proceed to system-level modelling. This involves creating an abstract model of the entire system to meet functional and performance requirements. The Design phase concludes with RTL design, refining the hardware implementation.

Then, the Synthesis and Test phases, generally occur simultaneously. During synthesis, RTL synthesis converts the RTL design into a gate-level netlist, followed by physical synthesis which transfer the netlist to physical representation, which places and routes the design to achieve the hardware’s physical layout. Meanwhile, testing ensures the design's correctness and functionality. Verification checks logical correctness, while system-level validation ensures the functionality and performance of the complete system align with the specifications.

Finally, after passing all Verification and Validation (V\&V) stages, the development transitions to the back-end \cite{ref13}. As our focus is solely on the front end, the back-end process is beyond the scope of this discussion.

Currently, algorithm and application engineers primarily define requirements and specifications for the hardware-level design, while hardware engineers handle the design, synthesis, and test stages. However, with the advent of Generative AI (Gen-AI), a new process has become possible. This process allows algorithm and application engineers to use Gen-AI to participate more deeply in the hardware development process. Increased involvement can enhance design quality, enhance development efficiency, and help identify and resolve potential issues earlier than the original design flow, ensuring effective and reliable hardware development.

As an extension of natural language, Gen-AI ranges from basic language constructs to advanced programming and hardware description languages (HDLs). This progression reflects the transition from simple language structures to complex programming methods and hardware specifications, bridging human communication and technical implementation. By enabling the seamless conversion of ideas into practical hardware designs, Gen-AI significantly accelerates the development process \cite{ref14}.

This technology has become a transformative force across multiple fields, profoundly impacting hardware development. Its potential to revolutionise hardware development lies in automating, streamlining, and injecting creativity into a field traditionally reliant on extensive manual effort and time.

The methodologies of Gen-AI are divided into two main approaches: interactive language learning and auxiliary code generation \cite{ref14}. Interactive language learning enables AI algorithms to improve their generation capabilities by learning from human-provided natural language descriptions. In contrast, auxiliary code generation uses AI to assist human engineers in coding tasks by providing suggestions and automating repetitive processes.

Current research on Gen-AI-driven HDL generation predominantly focuses on the component level \cite{ref15,ref16,ref17,ref18,ref19}. However, translating natural language into system-level HDL design remains relatively unexplored, offering a promising avenue for future research.

This research aims to accelerate hardware development by addressing the challenges of system-level HDL design using Gen-AI, thereby fostering innovation and enabling the development of more advanced, efficient hardware solutions. Detailed contributions are as follows:

\begin{itemize}
    \item \textbf{A Collaborative Development Pathway:} We propose a pathway that facilitates collaboration between application engineers, algorithm engineers and hardware engineers, enabling both to work within the hardware development process.
    \item \textbf{System-Level HDL Generation Tool:} We introduce the Natural-Level Synthesis (NLS) extension on Visual Studio Code (VS Code), a tool that uses Gen-AI to create system-level HDL designs from natural language descriptions. The tool’s name reflects its ability to develop complex designs from simple, intuitive inputs. This tool streamlines the design process and boosts productivity.
    \item \textbf{Benchmark:} We establish a benchmarking framework to systematically assess the tool's performance, which can be used to evaluate and compare similar tools developed either before or after this one.
    \item \textbf{Case Studies:} We conduct case studies using various Gen-AI models to demonstrate the effectiveness and versatility of the tool. Additionally, we address challenges associated with the tool and evaluate its performance.
\end{itemize}
The rest of the paper is organized as follows, Section II provides a comprehensive review of Gen-AI-based design tools for hardware design; Section III describes our methodology towards the collaborative development pathway, the system-level HDL generation approach and the benchmarking; Section IV includes the case studies to highlight the features of your new approach and benchmark. At last, we conclude the paper in Section V.

\section{Literature Review}
Several Gen-AI models have shown potential for application in hardware development. OpenAI’s OpenAI-o1, ChatGPT-4, Anthropic’s Claude 3.5 Sonnet, and Meta’s Llama 3.1 are among the leading models \cite{ref20,ref21,ref22,ref23}. These AI systems are adept at processing and generating human-like text, which could be directed towards interpreting complex technical specifications and generating the corresponding HDL code. Another type of Gen-AI model includes auxiliary coding tools, such as GitHub Copilot and Amazon Q Developer, which assist by suggesting code snippets and functions \cite{ref24,ref25}. While these tools can simplify HDL scripting, they are not well suited for writing HDL. They often require extensive manual adjustments to ensure accuracy and functionality \cite{ref14}.

The application of Gen-AI is not limited to language processing. Building on the foundation laid by HLS tools, recent research has explored the integration of Gen-AI to further simplify the hardware development process. This research involves developing tools capable of converting natural language directly into HDL. Such advancements could significantly enhance the accessibility and intuitiveness of hardware development, allowing even those without deep technical knowledge of HDL to participate in hardware development.

VeriGen is a Verilog code generation model fine-tuned on Verilog datasets from GitHub and textbooks \cite{ref15}. It outperforms models like GPT-3.5-turbo, particularly in generating syntactically correct Verilog code for complex scenarios, making it effective for hardware design automation.

VerilogEval consists of 156 problems taken from HDLBits, such as reversing the bit order of an 8-bit vector [7:0] or describing an "assign" function \cite{ref16,ref26}. All the problems are straightforward and focus on the component level.

RTLLM tests Gen-AI’s capability to generate functional hardware components such as arithmetic units like accumulators and different bit sizes adder, or logic units like right shifters and muxes \cite{ref17}. Chang's research also includes comparative analyses of components generated by various tools, including their product ChipGPT, ChatGPT, traditional HLS tools, and Chisel, to show differences in outputs such as decoders and adder-multi trees \cite{ref18}.

VeriAssist focuses on verifying and correcting Verilog code \cite{ref19}. It generates RTL code, tests it, and iteratively corrects issues using self-verification and self-correction, reducing the need for manual intervention and improving code quality.

All of these researches have focused on the component-level code generation, despite these advancements at the component level, there remains a significant gap in research concerning system-level designs using Gen-AI. This area, largely unexplored, offers immense potential for future research and could lead to major breakthroughs in how complex hardware systems are conceptualized and implemented using advanced AI technologies. The integration of Gen-AI into system-level design promises to expand the scope and efficiency of hardware development, paving the way for innovative solutions in the field of EDA.

\section{Methodology}

\subsection{A Collaborative Development Pathway}
Considering the existing hardware development process, we propose the pathways where Gen-AI enables both hardware engineers and algorithm or application engineers, to actively engage in the development stages. This division of labour creates a more specialised and efficient workflow.

In Fig. 2, the process begins with algorithm or application engineers and hardware engineers collaborating to define project requirements and specifications. During the Design phase, algorithm or application engineers leverage Gen-AI to generate HDL code based on predefined constraints. At the same time, hardware engineers develop high-level system models to simulate and analyse performance. Subsequently, algorithm engineers use Gen-AI to create detailed RTL designs.

\begin{figure}[!t]
\centering
\includegraphics[width=3.5in]{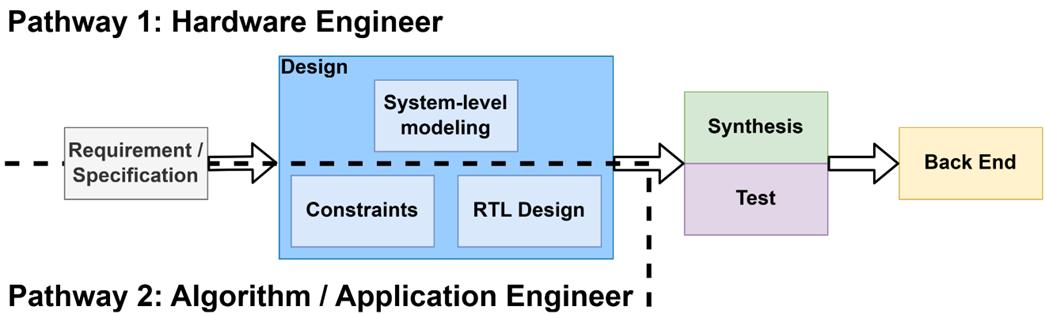}
\caption{New hardware development process with 2 pathways.}
\label{fig_2}
\end{figure}

In the Synthesis phase, hardware engineers translate the RTL design into a gate-level netlist, optimising for performance and efficiency \cite{ref13}. They also manage physical synthesis, addressing placement and routing constraints. During the Manufacturing stage, hardware engineers collaborate with manufacturing teams to ensure that production follows the physical layout, addressing any manufacturability and quality concerns. In the Test phase, both teams collaborate to verify hardware functionality and validate the complete system, ensuring it meets all requirements. Finally, the product is released, meeting quality standards and customer expectations.

Introducing Gen-AI into the hardware development process establishes two pathways that ultimately converge into a single cohesive project. The first pathway focuses on algorithm or application engineers leveraging Gen-AI to generate HDL code during the Design phase. The second pathway concentrates on hardware engineers focusing on system-level modelling and managing synthesis processes. These pathways converge at critical points, such as during the RTL Synthesis and Test stages, ensuring a collaborative and efficient hardware development process. This integration not only enhances productivity but also capitalises on the strengths of all algorithm, application and hardware engineers, resulting in a more robust and innovative hardware product.

\subsection{System-Level HDL Design Tool}
The NLS extension is illustrated in Fig. 3. This extension is a VS Code plugin that generates Verilog code using natural language prompts through OpenAI models. Leveraging the OpenAI API key, it automates design tasks directly within the VS Code environment. The workflow includes multiple stages of user-system interaction, facilitated by a sequence of prompts and commands.

\begin{figure}[!t]
\centering
\includegraphics[width=3.5in]{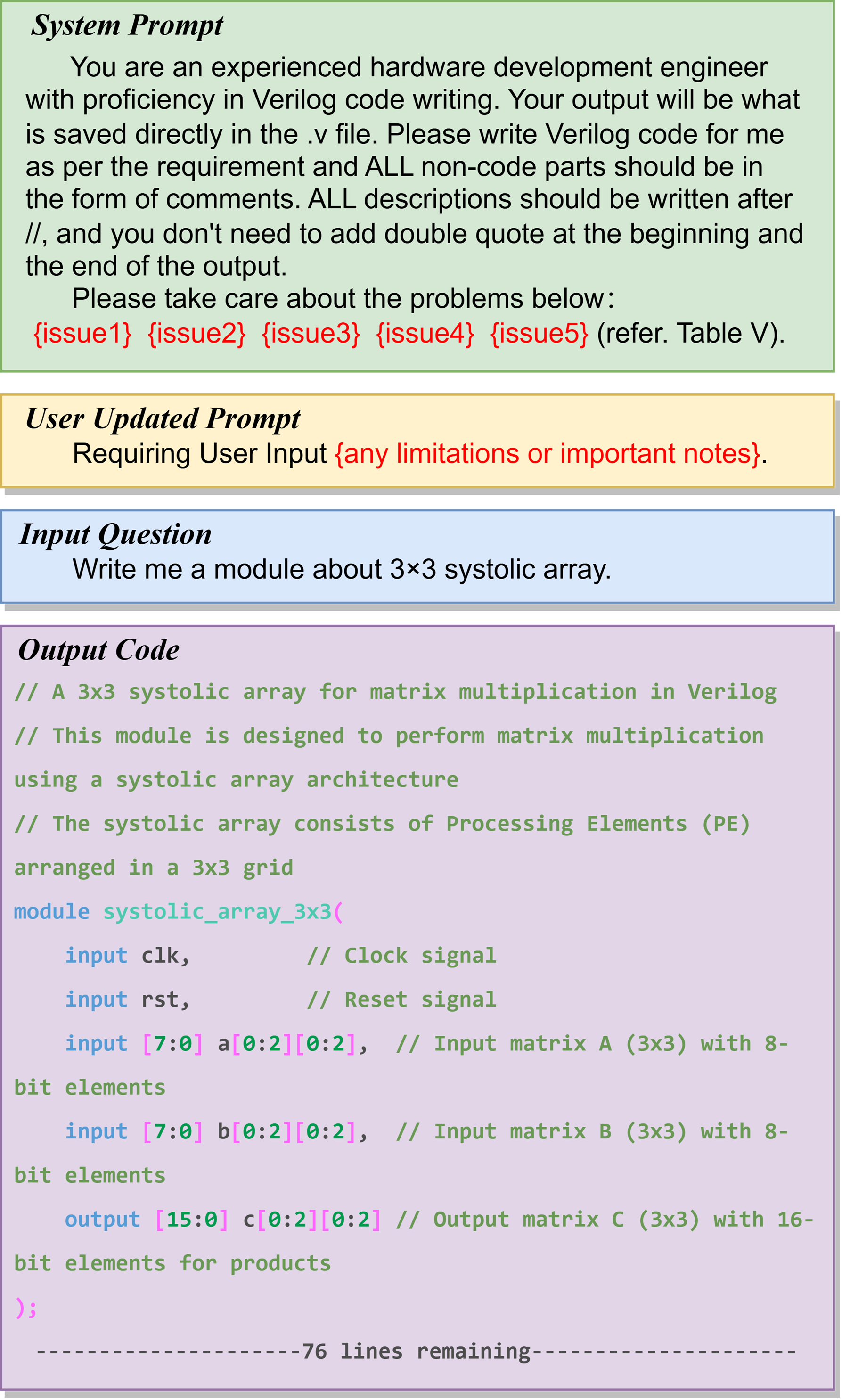}
\caption{Process of the NLS extension.}
\label{fig_3}
\end{figure}

In the example shown, the \textit{System Prompt} provides detailed instructions to the model, defining the expected format for code generation. The prompt specifies the model to generate Verilog code with comments for non-code sections and to avoid common issues outlined in the instructions. This guides the model to focus on these areas and avoid repeated errors, ensuring the output code is correct, clear, and maintainable.

The \textit{User Updated Prompt} enables modifications or notes to further refine requirements. The user can also add common issues to the System Prompt by using this section, enhancing model performance, and addressing specific problems effectively. This flexibility supports any real-time updates the user may require during development.

Finally, the \textit{Input Question} and \textit{Output Code} sections present a sample output generated by the extension. In this example, the GPT-4o model generates a Verilog module for a 3×3 systolic array, including detailed comments explaining the purpose of each input, output, and signal. This module is designed for matrix multiplication using a systolic array architecture. It includes inputs for two matrices, an output for the resulting matrix, and specified bit widths for precision.

The NLS extension includes several specific commands to streamline the development process.

\begin{itemize}
    \item \textbf{Add OpenAI API Key:} Users add their API key, required to access OpenAI models. Each new key replaces any previously entered key.
    \item \textbf{Select OpenAI Model:} Allow users to select a specific OpenAI model for code generation. They can first choose the model category (e.g., OpenAI-o1 category) and then a particular model within that category (e.g., OpenAI-o1-preview).
    \item \textbf{Generate Verilog Code:} This command initiates code generation once the API key and model selection commands are executed.
    \item \textbf{Updated Prompt:} This command allows users to input limitations or important notes and update them in the \textit{System Prompt}.
    \item \textbf{Package Verilog Code:} After code generation, this command compresses all .v files into a .zip archive for integration into other tools for simulation or synthesis.
\end{itemize}

After simulation or synthesis, users can use other VS Code extensions, such as Surfer, to display waveforms from .vcd files, aiding in visual verification of the Verilog code.

It is readily apparent that the code in the example is incorrect, as arrays cannot be declared directly within the port list. Therefore, it is crucial to summarise common issues and incorporate them into the \textit{System Prompt}. This will be explored further in the next case study section.

\subsection{Benchmark}
To evaluate the performance of different AI models in generating HDL code, we establish two benchmarks:

\subsubsection{Quality of Generated Hardware (QGH)}
Quality is measured using Performance, Power, and Area (PPA) metrics. Analyzing these factors reveals the efficiency and practicality of hardware designs. Comparing PPA results from different AI models helps identify those that produce better-optimized hardware.

\subsubsection{Required Design Efforts (RDE)}
Design efforts are evaluated by considering:

\begin{itemize}
    \item The length of the generated code (LoC).
    \item The character length of initial prompts given to the model (LoP).
    \item The length and number of adjustments needed from the first to the final prompt (LoA / NoA).
\end{itemize}

This indicates the effort needed to obtain satisfactory HDL code from each AI model. These benchmarks enable effective comparison of different AI models and aid in selecting the most suitable ones for generating HDL code.

\section{Case Study and Evaluation}
This section highlights several key problems of significant importance in the fields of AI and High-Performance Computing (HPC). However, engineers in these areas often lack expertise in hardware development. Additionally, we examined the performance of NLS on other HDL types.
\subsection{Case 1: The Discrete Poisson Equation (DPE)}
This case involves a Jacobi iteration function for solving the discrete Poisson equation, specifically to determine boundary values in a two-dimensional array, which is crucial for designing a parallel linear solver in HPC \cite{ref27}.

Initially, we used the ChatGPT-4o model to generate C code, converted it into a Vivado project by Vitis HLS  then synthesised it with Vivado for data comparison. We then generated Verilog code using multiple models and compared the synthesis results. Table I shows hardware resource usage for different Gen-AI models with NLS.

\begin{table*}[ht]
\centering
\caption{Case 1 - Hardware Resources Usage}
\label{tab:case1}
\begin{threeparttable}
\begin{tabular}{lllllllllll}
\toprule
\multirow{2}{*}{\textbf{Model}} & 
\multicolumn{7}{c}{\textbf{Hardware Resources}} &
\multirow{2}{*}{\textbf{Dynamic Power (W)}} & 
\multicolumn{2}{c}{\textbf{RDE}} \\ 
\cmidrule(lr){2-8}
\cmidrule(lr){10-11}
& \textit{LUTs} & \textit{Registers} & \textit{DSPs} & \textit{F7Muxes} & \textit{F8Muxes} & \textit{BRAM} & \textit{BUFGCTRL} & & \textit{LoP} & \textit{NoA}\\
\midrule
Hand Coding$^a$          & 7247 & 9197  & 14 & 12 & 1 & 2 & 1 & 147.72 & N/A & N/A\\
ChatGPT-4o$^a$         & 8989 & 6395  & 14 & 0 & 0 & 0 & 0 & 242.24 & 151 & 11 \\
OpenAI-o1-preview$^b$    & 8884 & 3580 & 0 & 445 & 2 & 0 & 1 & 207.22 & 151 & 13 \\
OpenAI-o1-mini$^b$     & 5102 & 3530  & 0 & 1111 & 92 & 0 & 1 & 51.65& 151 & 7 \\
Claude-3.5-sonnet$^b$    & 2995 & 2467  & 0 & 307 & 18 & 0 & 1 & 21.43 & 151 & 14 \\
\bottomrule
\end{tabular}
\begin{tablenotes}
\item $^a$~C code converted into a Vivado© project by Vitis HLS©, then synthesized using Vivado© 
\item $^b$~Verilog code synthesized using Vivado© 
\end{tablenotes}
\end{threeparttable}
\end{table*}

Both the OpenAI-o1-preview and OpenAI-o1-mini models generated Verilog hardware designs by interpreting natural language descriptions of the hand-coded C code. The Verilog code generated by the o1-preview model closely matches the computational flow and structure of the original C code, while the o1-mini model, though similar, captures these aspects with slightly less precision. According to Table I, the o1-preview model used approximately 4\% fewer LUTs and around 55\% fewer registers compared to the hand-coded C code. The o1-mini model used nearly 43\% fewer LUTs and 56\% fewer registers than the hand-coded C code, demonstrating notable efficiency in resource usage. Directly generating Verilog code, rather than using NLS to produce C code, results in more efficient resource usage, with both the o1-preview and o1-mini models consuming fewer resources than the ChatGPT-4o model. This highlights the resource-saving advantage of directly generating Verilog code.

In contrast, the Claude-3.5-sonnet and Llama-3.1 models produced less accurate results. Claude-3.5-sonnet created a functional Verilog design but significantly simplified the computational flow and convergence conditions of the original C code. The output of this model also contained bugs requiring manual correction. Resource-wise, Claude-3.5-sonnet consumed roughly 67\% fewer LUTs and around 69\% fewer registers than the hand-coded C code. The Llama-3.1 model, however, was unable to produce usable Verilog code, resulting in output that did not meet requirements.

For Required Design Efforts (RDE), the character length of initial prompts (LoP) remains consistent at 151 characters, while the number of adjustments (NoA) hovers around 10, reflecting the relatively small scale of this project.

\subsection{Case 2: LeNet-5}
LeNet is a pioneering convolutional neural network (CNN) architecture initially designed for handwritten digit recognition \cite{ref28}. Its structure, comprising multiple convolutional and pooling layers, has proven highly effective for feature extraction in computer vision \cite{ref29}.

Due to its relatively lightweight architecture, LeNet is an ideal candidate for hardware deployment, particularly on resource-constrained platforms like FPGAs and embedded devices \cite{ref30}. Deploying LeNet on hardware enables efficient real-time processing, reduces latency, and improves energy efficiency, making it suitable for applications requiring low power consumption and high-speed performance. 

\begin{figure}[!t]
\centering
\includegraphics[width=3.5in]{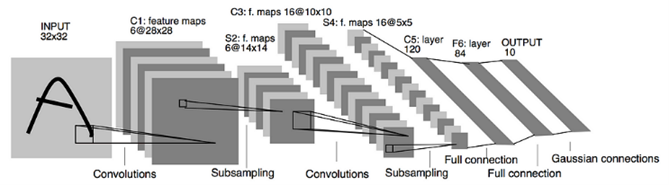}
\caption{LeNet-5 architecture \cite{ref28}.}
\label{fig_4}
\end{figure}

In this case, the LeNet-5 structure is deployed on a Zynq-7000 board using the OpenAI-o1-preview model. Table II compares hardware resource usage with other research. Percentages of higher usage are highlighted in green, while percentages of lower usage are highlighted in red.

\begin{table*}[ht]
\centering
\caption{Case 2 - Hardware Resources Usage}
\label{tab:case2}
\begin{threeparttable}
\begin{tabular}{lllllllll}
\toprule
& \multirow{2}{*}{\textbf{Target Board}} & \multicolumn{6}{c}{\textbf{Hardware Resources}} &
\multirow{2}{*}{\textbf{Power (W)}} \\ 
\cmidrule(lr){3-8} & & \textit{LUTs} & \textit{Registers} & \textit{DSPs} & \textit{BRAM} & \textit{F7Muxes} & \textit{F8Muxes} \\
\midrule
\textbf{NLS} & \textbf{Zynq-7000} & \textbf{34190} & \textbf{47254} & \textbf{10} & \textbf{0} & \textbf{3272} & \textbf{576} & \textbf{88.47} \\
\cite{ref30} & Zynq-7000 & 3092\textcolor{red}{(-90.96\%)} & 4095\textcolor{red}{(-91.34\%)} & 188\textcolor{teal}{(+1788.80\%)} & 44 & N/A & N/A & 0.374 \\
\cite{ref31} & Virtex-7 & 80175\textcolor{teal}{(+134.50\%)} & 46140\textcolor{red}{(-2.36\%)} & 83\textcolor{teal}{(+730.00\%)} & 0 & N/A & N/A & N/A \\
\cite{ref32} & Zynq-7000 & 39879\textcolor{teal}{(+16.64\%)} & 35399\textcolor{red}{(-25.09\%)} & 90\textcolor{teal}{(+800.00\%)} & 3 & N/A & N/A & N/A \\
\cite{ref33} & Artix-7 & 7986\textcolor{red}{(-76.64\%)}  & 3297\textcolor{red}{(-93.02\%)} & 200\textcolor{teal}{(+1900.00\%)} & N/A & N/A & N/A & 12\\
\bottomrule
\end{tabular}
\end{threeparttable}
\end{table*}

When generating Verilog code, NLS often follows a logic of loading all data at once before computation, resulting in significantly higher LUT resource usage. Despite this situation, NLS’s LUT usage is still more efficient than Zhou’s research \cite{ref31}, which used 134.5\% more LUT resources, and Ghaffari’s \cite{ref32}, which used 16.64\% more.

However, NLS’s approach to register management within Verilog is suboptimal, often resulting in an excessive or insufficient registers. This inefficiency led to increased register resource usage and, in certain instances, challenges in producing consistent results. Although NLS’s register usage was generally higher than in other studies, it was relatively comparable to Zhou’s usage \cite{ref31}.

Conversely, NLS performed exceptionally well in DSP usage, using only 10 DSPs—a substantial reduction compared to other studies, with a 94.7\% decrease relative to Ram’s research \cite{ref30} and a 95\% decrease relative to Mujawar’s \cite{ref33}. DSP resources on FPGAs are typically very limited, ranging from several hundreds to one or two thousand. They are more valuable than LUTs and registers. The reduced DSP usage may be due to bit incompatibilities. This leads to fewer DSPs being used but an increase in LUT usage. It represents a resource trade-off. For RDE, LoP is 2562 characters and NoA is 22.

\subsection{Case 3: picoMIPS processor using SystemVerilog}
NLS enables the construction of both Verilog and SystemVerilog projects. This case involves designing an 8-bit application-specific picoMIPS processor using System-Verilog \cite{ref34}. The goal was to efficiently implement an affine transformation algorithm.

We synthesised both the hand coding SystemVerilog code and the code generated by the OpenAI-o1-preview model in Vivado. The hardware resource usage and power consumption are shown in Table III. The meaning of the red and green highlights is consistent with the previous case.

Table III shows that SystemVerilog code generated by NLS exhibits similar issues to Verilog code. Both consume more LUTs and registers but use fewer DSPs. Since DSPs are a limited resource on FPGAs, reduced usage is beneficial as a resource trade-off. Additionally, lower DSP usage reduces dynamic power usage. For RDE, the LoP is 5637 characters due to the detailed descriptions, and the NoA is 15.

\begin{table*}[ht]
\centering
\caption{Case 3 - Hardware Resources Usage}
\label{tab:case3}
\begin{threeparttable}
\begin{tabular}{lllllllll}
\toprule
& \multicolumn{5}{c}{\textbf{Hardware Resources}} &
\multirow{2}{*}{\textbf{Dynamic Power (W)}} \\ 
\cmidrule(lr){2-6} & \textit{LUTs} & \textit{LUTRAMs} & \textit{Registers} & \textit{DSPs}  & \textit{BUFG} \\
\midrule
Hand Coding & 91 & 8 & 33 & 5 & 1 & 9.5 \\
\textbf{NLS} & \textbf{\textcolor{red}{230}} & \textbf{\textcolor{teal}{0}} & \textbf{\textcolor{red}{142}} & \textbf{\textcolor{teal}{0}} & \textbf{1} & \textbf{\textcolor{teal}{6.6}} \\
\bottomrule
\end{tabular}
\end{threeparttable}
\end{table*}

\subsection{Case 4: Adaptive Gradient Recurrent Neural Network}
This case involves using an adaptive gradient recurrent neural network (AGRNN) to effectively solve dynamic quadratic programming (DQP) problems with equational constraints. DQP is widely applied in many scientific and engineering fields, such as the robot control \cite{ref35,addRobot} and the power system \cite{ref36}. 

A general DQP problem with equality constraint is given as follows:
\begin{equation}
\begin{aligned}
&\underset{\boldsymbol{y}  }{\text{min}}  && \boldsymbol{y}^\text{T}(t) Q(t) \boldsymbol{y}(t)/2 + \boldsymbol{c}^\text{T}(t) \boldsymbol{y}(t) \\
&\text{s.t.} && W(t)\boldsymbol{y}(t) = \boldsymbol{z}(t),
\label{eq:1} 
\end{aligned}
\end{equation}
where time \( t \in [0, +\infty) \); \( \boldsymbol{y}(t) \in \mathbb{R}^n \) is unknown and needs to be solved in real time; the time-varying coefficient matrix \( Q(t) \in \mathbb{R}^{n \times n} \) is symmetric positive-definite, and \( W(t) \in \mathbb{R}^{m \times n} \) with \(m < n\) is of full row rank; the coefficient vector \( \boldsymbol{c}(t) \in \mathbb{R}^n \), and \( \boldsymbol{z}(t) \in \mathbb{R}^m \). 
Besides, the superscript \(^\text{T}\) represents the transpose operation.
To solve DQP problem (\ref{eq:1}), the equality constraint is handled with the help of the Lagrangian method\cite{addLagrangian}, so that DQP problem (1) can be converted into the form of the equation as $\ A(t)\boldsymbol{x}(t) = \boldsymbol{b}(t), $
where
\[
\begin{aligned}
    A(t) &= \begin{bmatrix} Q(t) & W^\text{T}(t) \\ W(t) & \boldsymbol{0}_{m \times m} \end{bmatrix} \in \mathbb{R}^{(n+m) \times (n+m)}, \\
    \boldsymbol{x}(t) &= \begin{bmatrix} \boldsymbol{y}(t) \\ \boldsymbol{\lambda}(t) \end{bmatrix} \in \mathbb{R}^{n+m},\; 
    \boldsymbol{b}(t) = \begin{bmatrix} -\boldsymbol{c}(t) \\ \boldsymbol{z}(t) \end{bmatrix} \in \mathbb{R}^{n+m},
\end{aligned}
\]
and $\boldsymbol{\lambda}(t) \in \mathbb{R}^m$ is the Lagrange multiplier vector. Next, an error function can be set to $\boldsymbol{e}(t) = A(t)\boldsymbol{x}(t) - \boldsymbol{b}(t)$. The goal is to make $\boldsymbol{e}(t)$ tend to $\boldsymbol{0}$ to approximate the theoretical solution.

Our previous research introduced an AGRNN model to solve DQP problems. The model was implemented on FPGA. The efficiency of AGRNN for solving DQP problems is significantly improved by the FPGA implementation. The presented AGRNN model is as in:
\begin{equation}
\label{eq1}
\dot{ \boldsymbol{x}}(t) = -\,k(t)\,A^\text{T}(t)\,\bigl(A(t)\, \boldsymbol{x}(t)\;-\;\boldsymbol{b}(t)\bigr),
\end{equation}
with \(\displaystyle k(t) \;=\; \beta \,\cdot\,
\frac{\bigl|(A(t)\,\boldsymbol{x}(t)\;-\;\boldsymbol{b}(t))^\text{T}\,\bigl(\dot{A}(t)\,\boldsymbol{x}(t)\;-\;\dot{\boldsymbol{b}}(t)\bigr)\bigr|}
     {\bigl\|\,A(t)\,\bigl(A(t)\,\boldsymbol{x}(t)\;-\;\boldsymbol{b}(t)\bigr)\bigr\|_{2}^{2}},\)

\noindent where \(\beta > 1\) denotes a constant; and \(\|\cdot\|_2\) denotes the \(L_2\)-norm of a vector.

Through the solution via the AGRNN model, the error $\boldsymbol{e}(t)$ can be made to gradually approach $\boldsymbol{0}$, which is equivalent to solving the DQP Problem (1) in real time.

This functionality is reconstructed using the GPT-4o model, with total hardware resource usage and specific module details presented in Table IV. The meaning of the red and green highlights is consistent with the previous case. For RDE, LoP is 1104 characters and NoA is 93. 

The complexity of calculating adaptive coefficients (AC) and generating the trigonometric function (TF) module prevents NLS from accomplishing the related code generation tasks. Consequently, a fixed scaling factor replaces the adaptive coefficient k(t) in the implementation, reverting the solver to a traditional gradient-based neural network. Simultaneously, trigonometric functions involved in time-varying parameters, such as sin(t), are replaced with the time-varying parameter t (TVP).

Compared to hand-coded Verilog, NLS-generated version used 15\% more LUTs and 9\% more DSPs but saved 10\% in registers after excluding differing elements. Focusing on each module, NLS-generated Verilog consumed approximately 30\% more resources than hand-coded Verilog, comparable to current HLS methods. However, using NLS significantly reduced the time and effort required for development and simulation. This case employed the ChatGPT-4o model. We believe newer models, such as OpenAI-o1-preview, could provide better resource efficiency.

\begin{table*}[htbp]
\centering
\caption{Case 4 - Hardware Resources Usage}
\label{tab:case4}
\begin{tabular}{lllll}
\toprule
& \multicolumn{4}{c}{\textbf{Hardware Resources}} \\
\cmidrule(lr){2-5} & \textit{LUTs} & \textit{Registers} & \textit{DSPs} & \textit{BUFGCTRL} \\
\midrule
\textbf{NLS}                          & \textbf{3116}         & \textbf{5676}        & \textbf{252}          & \textbf{1} \\
Hand Coding                  & 19053        & 13732       & 687          & 2 \\
\textbf{NLS -- excl. AC TF, w. TVP}             & \textbf{2819 \textcolor{teal}{(\,+15.30\%)}}  & \textbf{4811 \textcolor{red}{(\,-9.60\%)}}  & \textbf{216 \textcolor{teal}{(\,+9.09\%)}}  & \textbf{1 \textcolor{red}{(\,-50\%)}} \\
Hand Coding    & 2445         & 5322        & 198          & 2 \\
\textbf{NLS -- multiplier}            & \textbf{59}           & \textbf{116 \textcolor{teal}{(\,+73.13\%)}}  & \textbf{9}            & \textbf{0} \\
Hand Coding -- multiplier    & 59           & 67          & 9            & 0 \\
\textbf{NLS -- AXB}                   & \textbf{994 \textcolor{teal}{(\,+34.32\%)}}  & \textbf{1482 \textcolor{teal}{(\,+37.86\%)}}  & \textbf{81 \textcolor{teal}{(\,+12.5\%)}}   & \textbf{0} \\
Hand Coding -- AXB           & 740          & 1075        & 72           & 0 \\
\textbf{NLS -- x\_dot\_update}        & \textbf{1010 \textcolor{red}{(\,-13.82\%)}} & \textbf{2028 \textcolor{teal}{(\,+18.25\%)}}  & \textbf{108 \textcolor{teal}{(\,+9.09\%)}}  & \textbf{0} \\
Hand Coding -- x\_dot\_update & 1172         & 1715        & 99           & 0 \\
\bottomrule
\end{tabular}
\end{table*}

\subsection{Issues}
In the previous case studies, several common issues with NLS were identified. These issues primarily stem from the way models interpret natural language and generate HDL code. The issues are categorized in Table V.

\begin{table*}[htbp]
\centering
\caption{Common Issues and Solutions (Refer. Fig. 3)}
\label{tab:issues_solutions}
\begin{tabular}{lp{6cm}p{6cm}}
\toprule
\textbf{Issue} & \textbf{Description} & \textbf{Updated Prompt} \\
\midrule
1. Incorrect Register Usage &
Registers are often not managed appropriately, leading to either excessive or insufficient allocation. &
Ensure registers are managed correctly for optimal resource usage. \\
\midrule
2. Always Block Issues &
Declaring variables improperly within ``always'' blocks often leads to simulation errors and unknown output values. &
Declare variables globally instead of within ``always'' blocks to prevent simulation errors. \\
\midrule
3. SystemVerilog and Verilog Compatibility &
SystemVerilog syntax in Verilog designs can cause compatibility issues. &
Avoid using SystemVerilog syntax, such as array parameters or ``typedef'' in Verilog designs. \\
\midrule
4. Logic Errors &
Incorrect logic in ``always'' blocks or state machines leads to incorrect behaviour. &
Write accurate logic for ``always'' blocks and state machines, ensuring correct transitions. \\
\midrule
5. Fixed-Point Arithmetic Challenges &
Fixed-point numbers may be generated incorrectly. &
Model fixed-point arithmetic in Python before translating it to Verilog. \\
\bottomrule
\end{tabular}
\end{table*}

We strongly recommend updating the \textit{System Prompt} with these common issues. This will assist Gen-AI models in generating more accurate and maintainable HDL code.

\section{Future Work}
Our future research will concentrate on the following areas:

\textbf{Dataset Preparation and Model Training: }We aim to create a dedicated dataset to train a customized model for generating system-level HDL designs. The dataset will include diverse system-level HDL examples to enhance the model's capability to generate optimized and functional designs. The objective is to enhance the precision of natural-level synthesis and expand the applicability of our approach.

\textbf{System Partitioning Considerations: }Current research primarily focuses on the functional aspects of system-level HDL design. However, system partitioning, an important non-functional consideration, remains unexplored. Our future work aims to address this gap by investigating AI-driven system partitioning. System partitioning is a crucial non-functional factor that significantly influences design quality and efficiency. We will integrate an AI-based partitioning tool to streamline system integration further, reducing the time and effort required for design iterations.

\textbf{Extension Features: }Our NLS tool will be enhanced with additional features to improve user experience. These features include compatibility with locally trained models, enabling users to choose among different Gen-AI models, generating a wider variety of HDL, and the ability to modify previously generated code. These enhancements will make the tool more versatile, improve its efficiency, and broaden its potential applications.

\section{Conclusion}
This study introduces a novel tool, Natural-Level Synthesis (NLS), which employs generative AI to transform natural language descriptions into system-level HDL code. Developing the NLS extension demonstrates the potential for engineers across disciplines to participate in the hardware design process, simplifying HDL code generation and bridging the gap between algorithms and hardware engineering.

The NLS extension facilitates the automatic generation of HDL code while offering an intuitive environment for code development, significantly reducing hardware design time. However, this advantage comes with increased resource usage, which must be managed effectively. Results from benchmarks and case studies reveal that generative AI reduces development time while maintaining functionality, demonstrating the feasibility of this approach for practical applications. 

{\appendix[The GitHub Links for the Threads and Codes]}
Threads for cases:
\par \url{https://github.com/k-yang11/NLS}

Codes for NLS:
\par \url{https://github.com/k-yang11/NLS_Extension}

\newpage

\section{Biography Section}
\vspace{11pt}

\begin{IEEEbiography}[{\includegraphics[width=1in,height=1.25in,clip,keepaspectratio]{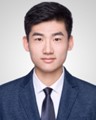}}]{Kaiyuan Yang}
is a current PhD candidate at University of Sheffield, United Kingdom. He received the BEng degree in Electronic and Computer Engineering from University of Sheffield in 2022. His research focuses on hardware acceleration for deep learning and AI-EDA for hardware design.
\end{IEEEbiography}

\begin{IEEEbiography}[{\includegraphics[width=1in,height=1.25in,clip,keepaspectratio]{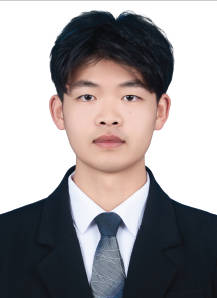}}]{Huang Ouyang}
received the B.E. degree in computer science and technology from JiShou University, JiShou, China, in 2023. He is currently pursuing an M.S. degree in computer application technology at Lanzhou University, Lanzhou, China.
His main research interests include neural networks, hardware acceleration, and field programmable gate arrays.
\end{IEEEbiography}

\begin{IEEEbiography}[{\includegraphics[width=1in,height=1.25in,clip,keepaspectratio]{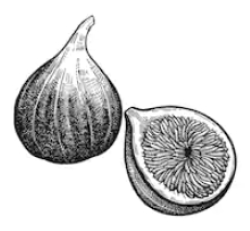}}]{Xinyi Wang}
is currently a PhD student at the School of EEE at the University of Sheffield. She received her MSc degree from the University of Edinburgh and BEng degree from the University of Sheffield. 
\end{IEEEbiography}

\begin{IEEEbiography}[{\includegraphics[width=1in,height=1.25in,clip,keepaspectratio]{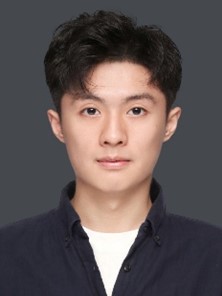}}]{Bingjie Lu}
is currently a postgraduate student at the University of Sheffield, United Kingdom. He received his BEng degree in Electronic and Electrical Engineering from University of Sheffield in 2023. His main research interests focus on FPGA hardware design, SystemVerilog-based digital design and sensor circuit design.
\end{IEEEbiography}

\begin{IEEEbiography}[{\includegraphics[width=1in,height=1.25in,clip,keepaspectratio]{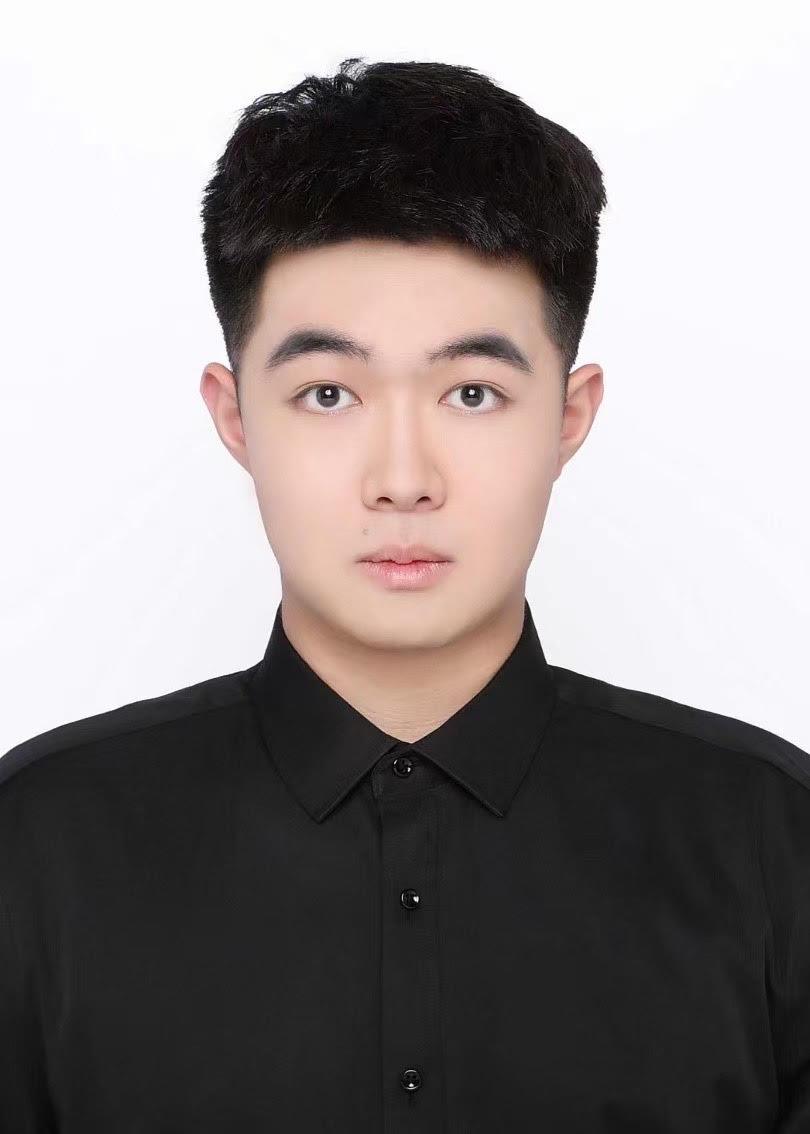}}]{Yanbo Wang}
received his MSc degree from School of EEE, the University of Sheffield. 
\end{IEEEbiography}

\begin{IEEEbiography}[{\includegraphics[width=1in,height=1.25in,clip,keepaspectratio]{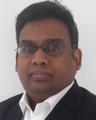}}]{Charith Abhayaratne}
(Member, IEEE) received the B.E. degree in electrical and electronic engineering from University of Adelaide, Australia, in 1998, and the Ph.D. degree in electronic and electrical engineering from the University of Bath, U.K., in 2002. He is currently a Senior Lecturer with the School of Electronic and Electrical Engineering, University of Sheffield, U.K. He has published over 90 peer-reviewed papers in leading journals, conferences, and book editions. His research interests include visual content analysis, visual content security, machine learning, and multidimensional signal processing. He was a recipient of the European Research Consortium for Informatics and Mathematics (ERCIM) Postdoctoral Fellowship to carry out research from the Centre of Mathematics and Computer Science (CWI), The Netherlands, from 2002 to 2004, and the National Research Institute for Computer Science and Control (INRIA), Sophia Antipolis, France. He currently serves as an Associate Editor for IEEE TRANSACTIONS ON IMAGE PROCESSING, IEEE ACCESS, and Journal of Information Security and Applications (JISA) (Elsevier).
\end{IEEEbiography}

\begin{IEEEbiography}[{\includegraphics[width=1in,height=1.25in,clip,keepaspectratio]{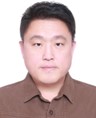}}]{Sizhao Li}
(Member, IEEE) received the B.S. degree from Jilin University, Changchun, China, in 2007, the M.S. degree from the University of Science and Technology of China, Hefei, China, in 2011, and the Ph.D. degree in electrical engineering from Xiamen University, Xiamen, Fujian, China, in 2018. In 2018, he joined the College of Computer Science and Technology, Harbin Engineering University, Harbin, Heilongjiang, China, as an Associate Professor. He was a Visiting Scholar with the University of Illinois at Urbana–Champaign, Champaign, IL, USA.

He is also the Vice Director of the Intelligent Computing and Industrial Inter- net Security Center, College of Computer Science and Technology, Harbin Engineering University. His research interests include high-performance par- allel computing, computer architecture, and design of artificial intelligence systems.

\end{IEEEbiography}

\begin{IEEEbiography}[{\includegraphics[width=1in,height=1.25in,clip,keepaspectratio]{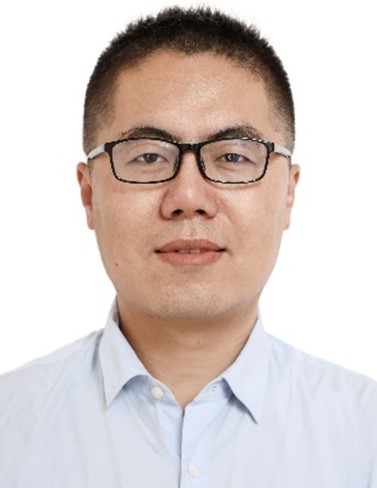}}]{Long Jin}
(Senior Member, IEEE) received the B.E. degree in automation and the Ph.D. degree in information and communication engineering from Sun Yat-sen University, Guangzhou, China, in 2011 and 2016, respectively.
		
He underwent postdoctoral training with the Department of Computing, The Hong Kong Polytechnic University, Hong Kong, from 2016 to 2017. In 2017, he was a Professor of Computer Science and Engineering with the School of Information Science and Engineering, Lanzhou University, Lanzhou, China. From 2023 to 2024, he served as a Visiting Professor with The City University of Hong Kong, Hong Kong. His current research interests include neural networks, optimization, intelligent computing, and robotics.
		
Prof. Jin currently serves as an Associate Editor for several journals such as IEEE Transactions on Intelligent Vehicles, IEEE Transactions on Industrial Electronics, and Neural Networks.
\end{IEEEbiography}

\begin{IEEEbiography}[{\includegraphics[width=1in,height=1.25in,clip,keepaspectratio]{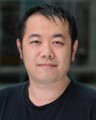}}]{Tiantai Deng}
received his PhD from Queen’s University Belfast, He is currently a lecturer at the University of Sheffield. Prior to his career as an academic, he was a senior engineer at HiSilicon, Huawei. His main research focus is on hardware acceleration for image processing, deep learning and high-level design environments.

\end{IEEEbiography}

\vspace{11pt}

\vfill

\end{document}